\documentstyle[12pt,aps,prc,epsfig,preprint]{revtex} 
\tightenlines 
\begin{document} 
\begin{titlepage} 

\title{Faddeev-Yakubovsky calculations for light 
$\Lambda\Lambda$ hypernuclei} 
\author{I.N. Filikhin$^a$, A. Gal$^a$ \\ 
$^a$Racah Institute of Physics, The Hebrew University, 
Jerusalem 91904, Israel} 
\maketitle 

\begin{abstract}
New Faddeev-Yakubovsky calculations are reported for 
$_{\Lambda\Lambda}^{~~6}$He and $_{\Lambda\Lambda}^{~10}$Be 
in terms of clusters of $\alpha$'s and $\Lambda$'s, 
using $\Lambda\Lambda$ $s$-wave potentials motivated by 
several of the Nijmegen model interactions. 
The self consistency of the $\Lambda\Lambda$ hypernuclear 
world data for these species is discussed. 
The newly reported $_{\Lambda\Lambda}^{~~6}$He event is found 
to be compatible with $\Lambda\Lambda$ interaction strengths 
provided by the Nijmegen soft-core one-boson-exchange model 
NSC97. Faddeev calculations for $_{\Lambda\Lambda}^{~~5}$H 
and $_{\Lambda\Lambda}^{~~5}$He suggest that these 
$\Lambda\Lambda$ hypernuclei are stable against emitting 
$\Lambda$'s for any (essentially attractive) $\Lambda\Lambda$ 
interaction, whereas calcualtions for $_{\Lambda\Lambda}^{~~4}$H 
do not allow a clear-cut conclusion whether or not it is 
particle stable. 
\newline 
\newline 
\newline 
$PACS$: 21.80.+a, 11.80.Jy, 21.45.+v 
\newline 
\newline 
\newline 
\newline 
Corresponding author: Avraham Gal 
\newline 
Tel: +972 2 6584930, Fax: +972 2 5611519 
\newline 
E mail: avragal@vms.huji.ac.il 
\newline 
\newline 
\newline 
\newline 
\centerline{(\today )}
\end{abstract} 
\end{titlepage}

\section{Introduction} 
\label{sec:int} 

Data on strangeness $S=-2$ hypernuclear systems is rather scarce, 
and no data exist for systems with higher strangeness content of 
hyperons ($Y$). Multistrange hadronic matter in finite systems and 
in bulk is predicted on general grounds to be stable, up to strangeness 
violating weak decays (see Ref. \cite{SBG00} for a recent review). 
Until recently only three candidates existed for $\Lambda\Lambda$ 
hypernuclei which fit events seen in emulsion experiments 
\cite{Dan63,Pro66,Aok91}. The $\Lambda\Lambda$ binding energies 
deduced from these events indicated that the $\Lambda\Lambda$ 
interaction is strongly attractive in the $^{1}S_0$ channel 
\cite{DMGD91,YTI91}, although it had been realized \cite{BUC84,WTB86} 
that the binding energies of the two older events, 
$_{\Lambda\Lambda}^{~~6}$He and $_{\Lambda\Lambda}^{~10}$Be, 
are inconsistent with each other. 
This outlook is perhaps undergoing an important change following the 
very recent report from the KEK hybrid-emulsion experiment E373 
on a new candidate \cite{Tak01} for $_{\Lambda\Lambda}^{~~6}$He, with 
binding energy substantially lower than that deduced from the older 
event \cite{Pro66}. Furthermore, there are also indications from the 
AGS experiment E906 for the production of light $\Lambda\Lambda$ 
hypernuclei \cite{Ahn01}, perhaps as light even as 
$_{\Lambda\Lambda}^{~~4}$H, in the $(K^-,K^+)$ reaction on $^9$Be. 

The study of multistrange systems can provide rather stringent 
tests of microscopic models for the baryon-baryon interaction. 
Over the years, the Nijmegen group has constructed 
a number of one-boson-exchange (OBE) models for the baryon-baryon 
interaction ($NN$, $\Lambda N - \Sigma N$, and 
$\Xi N -\Lambda\Lambda - \Sigma\Sigma -\Lambda\Sigma$) 
using SU(3)-flavor symmetry to relate baryon-baryon-meson coupling 
constants and phenomenological hard or soft cores at short distances 
(for a recent review, see Ref. \cite{Rij01}). 
In addition, the J\"ulich group has constructed OBE models 
for the $YN$ interaction along the lines of the Bonn Model for 
the $NN$ interaction, using SU(6) symmetry to relate coupling constants 
and short-range form factors \cite{RHS94}. The SU(6) quark model has 
also been used to derive baryon-baryon interactions within a unified 
framework of a $(3q)-(3q)$ resonating group method, augmented by a few 
effective meson exchange potentials of scalar and pseudoscalar meson 
nonets directly coupled to quarks \cite{FNS96}. In all of these 
rather different baryon-baryon interaction models only 35 $YN$ 
low-energy, generally imprecise data points serve for the purpose 
of steering phenomenologically the extrapolation from the $NN$ sector, 
which relies on thousands of data points, into the strange 
$YN$ and $YY$ sectors. It is therefore of utmost importance to confront 
these models with the new $\Lambda\Lambda$ hypernuclear data in order 
to provide meaningful constraints on the extrapolation to $S = -2$ 
and beyond. 

In this paper we report on Faddeev-Yakubovsky three-body and four-body 
calculations of light $\Lambda\Lambda$ hypernuclei, using 
$\Lambda\Lambda$ $s$-wave potentials which simulate the low-energy 
$s$-wave scattering parameters produced by several Nijmegen 
OBE models. Our own work is compared to other recent variational 
calculations \cite{HKM97,NSF00,YNa00}. The advantage of the 
Faddeev-Yakubovsky method is that it accounts properly for ${\it all}$ 
the relevant rearrangement channels. The purpose of our calculations 
is twofold: to check the self consistency of the data, particularly 
for $_{\Lambda\Lambda}^{~~6}$He and $_{\Lambda\Lambda}^{~10}$Be 
\cite{DDF89} which are treated here as clusters of $\alpha$'s and 
$\Lambda$'s; and to find out which of the recent Nijmegen Soft-Core 
(NSC97) models \cite{RSY99,SRi99}, or the Extended Soft-Core (ESC00) 
interaction model \cite{Rij01}, is the most appropriate one 
for describing well $\Lambda\Lambda$ hypernuclei. 
The paper is organized as follows. The Faddeev-Yakubovsky approach, 
its restriction to $s$ waves for light hypernuclei, and some computational 
aspects are described briefly in Sect. \ref{sec:meth}. We specify the 
input potentials, fitting as much as possible known binding energies of 
given hypernuclear subsystems. The results of our Faddeev-Yakubovsky 
calculations for $_{\Lambda\Lambda}^{~~6}$He and 
$_{\Lambda\Lambda}^{~10}$Be in terms of clusters of $\alpha$'s and 
$\Lambda$'s, using $\Lambda\Lambda$ $s$-wave potentials motivated by 
several Nijmegen model interactions, are reported and discussed 
in Sect. \ref{sec:res}. Faddeev calculations for $_{\Lambda\Lambda}^{~~5}$H 
and $_{\Lambda\Lambda}^{~~5}$He are also reported. A preliminary, partial 
report of these results has just appeared as a Rapid Communication 
\cite{FGa02}. In the Summary section we briefly discuss the implications 
of our work on the stability of $_{\Lambda\Lambda}^{~~4}$H, 
for which our Faddeev-Yakubovsky calculations at present are in progress.

\section{Methodology} 
\label{sec:meth} 

\subsection{Faddeev equations} 

The bound states of the three-body systems considered in this work 
are obtained by solving the differential Faddeev equations \cite{MFa93} 
\begin{equation} 
\label{eq:Fad}
(H_{0}+V^{c}_\beta+ V^s_\beta-E)U_\beta=-V^{s}_\beta\sum^3_{\gamma=1,
\gamma \neq\beta}U_\gamma \;\;\;, \;\;\;\;\;\; \beta=1,2,3 \;\;\;,
\end{equation}
where $V^c_\beta$ and $V^{s}_\beta$ are Coulomb and short-range pairwise 
interactions respectively in the channel $\beta$, $H_0$ is the kinetic 
energy operator, $E$ is the total energy and the wavefunction of the 
three-body system is given as a sum over the three Faddeev components, 
$\Psi =\sum^3_{\beta=1 }U_\beta$, corresponding to the two-body breakup 
channals. The Faddeev components are functions of spin-isospin variables 
and of relative Jacobi coordinates. 
When two constituents of the three-body system are identical, for example 
$\Lambda$ hyperons in $_{\Lambda\Lambda}^{~~6}$He ($\alpha \Lambda \Lambda$) 
generically of the form $C \Lambda \Lambda$ ($C$ = core), 
the coupled set of Faddeev equations simplifies as follows: 
\begin{equation} 
\label{eq:LLC}
(H_0+V_{\Lambda\Lambda}-E)U=-V_{\Lambda\Lambda}(W-P_{23}W),\;\;\; 
(H_0+V_{\Lambda C}-E)W=-V_{\Lambda C}(U-P_{23}W), 
\end{equation} 
where $P_{23}$ is the permutation operator for the $\Lambda$ hyperons 
(particles 2,3). $U$ is the Faddeev component corresponding to the 
rearrangement channel $\alpha-(\Lambda\Lambda)$ and $W$ corresponds 
to the rearrangement channel $(\alpha\Lambda)-\Lambda$: 
\begin{equation} 
\label{eq:PsiLLC} 
\Psi_{\alpha-(\Lambda\Lambda)} = U \;, \;\;\; 
\Psi_{(\alpha\Lambda)-\Lambda} = (1 - P_{23})W \;; \;\;\;\;\; 
\Psi = \Psi_{\alpha-(\Lambda\Lambda)} + \Psi_{(\alpha\Lambda)-\Lambda} \;. 
\end{equation} 
The probability weights for the various rearrangement channels $\beta$ in 
the total wavefunction $\Psi$ of $_{\Lambda\Lambda}^{~~6}$He are defined by 
\begin{equation} 
\label{eq:prob} 
P_{\beta} =  \frac{<\Psi,\Psi_{\beta}>}{<\Psi,\Psi>} \;. 
\end{equation} 
For $_{\Lambda}^9$Be, here considered as a $\Lambda \alpha \alpha$ system, 
a similar simplification occurs, but where the minus sign in 
Eqs. (\ref{eq:LLC}) in front of $P_{23}$ arising from the 
Fermi-Dirac statistics for the $\Lambda$'s is replaced by a plus sign 
owing to the Bose-Einstein statistics for the $\alpha$'s, and $V^c$ is 
added on the l.h.s. of the first Faddeev equation.  
These Faddeev equations, for $_{\Lambda}^9$Be(${\frac{1}{2}}^+$) and 
$^{~~6}_{\Lambda\Lambda}$He($0^+$) ground states, were solved in 
Ref. \cite{FYa00b} using an $s$-wave approximation in which only 
the $\ell_{\alpha}=0$, $\lambda_{\alpha}=0$ partial waves were retained 
for the Jacobi-coordinate $\bf x_{\alpha}$, $\bf y_{\alpha}$ degrees 
of freedom, respectively, where 
\begin{equation} 
\label{eq:coor} 
{\bf x}_\alpha={\bf r}_\beta - {\bf r}_\gamma , \qquad 
{\bf y}_\alpha=\frac{m_\beta{\bf r}_\beta+m_\gamma{\bf r}_\gamma} 
{m_\beta+m_\gamma} -{\bf r}_\alpha, \qquad \beta > \gamma \neq \alpha, 
\quad \alpha,\beta,\gamma=1,2,3, 
\end{equation} 
where $m_\alpha$ and ${\bf r}_\alpha$, $\alpha$=1,2,3, are the mass 
and the position vector of the $\alpha$-th constituent, respectively. 
(The total orbital angular momentum is given here by 
$\bf L = {\vec \ell}_{\alpha}+{\vec \lambda}_{\alpha}$). 
The present calculation extends and updates the previous one. 
We comment in Sections \ref{sec:res} and \ref{sec:sum} on the 
validity of the $s$-wave approximation. 

Other three-body systems studied in the present work are the 
isodoublet $_{\Lambda\Lambda}^{~~5}$H, $_{\Lambda\Lambda}^{~~5}$He 
charge-symmetric hypernuclei, here considered as $^3$H$\Lambda\Lambda$ 
and $^3$He$\Lambda\Lambda$ respectively. For the ground state 
($\frac12^+$) of this system, the $s$-wave Faddeev equations assume 
the form 
$$
(H_0^u(x_1,y_1)+v^s_{\Lambda\Lambda}(x_1)-E){\cal U}(x_1,y_1)=
-v^s_{\Lambda\Lambda}(x_1) \int\limits_{-1}^1 dv_1\frac{x_1y_1}{x'_1y'_1}
A{\cal W}(x'_1,y'_1),
$$
\begin{equation} 
\label{eq:fad} 
(H_0^w(x_2,y_2)+V_{\Lambda C}(x_2)-E){\cal W}(x_2,y_2)=-\frac12 
V_{\Lambda C}(x_2)\left[\int\limits_{-1}^1dv_2\frac{x_2y_2}{x'_2y'_2}
A^T{\cal U}(x'_2,y'_2)  \right.
\end{equation}
$$
\left. +\int\limits_{-1}^1dv_2\frac{x_2y_2}{x''_2y''_2} B{\cal W}(x''_2,y''_2) \right] ,
$$
where $v^s_{\Lambda\Lambda}(x)$ is the singlet $\Lambda\Lambda$ potential,
$V_{\Lambda C}=diag\{v^s_{\Lambda C},v^t_{\Lambda C}\}$ with 
$v^s_{\Lambda C}$ and $v^t_{\Lambda C}$ the singlet and triplet 
$\Lambda C$ interaction potentials, respectively, 
$$
H_0^{u}(x,y)=-\frac{\hbar^2}{2M_1}\partial^2_y
-\frac{\hbar^2}{2\mu_1}\partial^2_x, \qquad
H_0^{w}(x,y)=-\frac{\hbar^2}{2M_2}\partial^2_y
-\frac{\hbar^2}{2\mu_2}\partial^2_x,
$$
$$
A=(-\frac12,-\frac{\sqrt3}2), \qquad
B=
\left( \begin{array}{cc}
-\frac12& \frac{\sqrt3}2 \\
\frac{\sqrt3}2& \frac12\\
\end{array} \right) ,
$$
$$
{\cal W}(x,y)=({\cal W}^s(x,y),{\cal W}^t(x,y))^T.
$$
The appropriate reduced masses are given by 
$$
M_1=2\frac{m_Cm_{\Lambda}}{m_C+2m_{\Lambda}}, \qquad 
M_2=\frac{m_{\Lambda}(m_C+m_{\Lambda})}{m_C+2m_{\Lambda}}, \qquad 
\mu_1=\frac{m_{\Lambda}}{2}, \qquad 
\mu_2=\frac{m_Cm_{\Lambda}}{m_C+m_{\Lambda}}.
$$
The appropriate transformation of coordinates is given by 
$$
x'_1=((\frac{\mu_1}{m_{\Lambda}}x_1)^2+{y_1}^2+2\frac{\mu_1}{m_{\Lambda}}x_1y_1v_1)^{1/2},
\quad y'_1=\frac{m_C}{m_C+m_{\Lambda}}((\frac{m_{\Lambda}}{M_1}x_1)^2+{y_1}^2-
2\frac{m_{\Lambda}}{M_1}x_1y_1v_1)^{1/2},
$$
$$
x''_1=((\frac{\mu_1}{m_{\Lambda}}x_1)^2+{y_1}^2-2\frac{\mu_1}{m_{\Lambda}}x_1y_1v_1)^{1/2},
\quad y''_1=\frac{m_C}{m_C+m_{\Lambda}}((\frac{m_{\Lambda}}{M_1}x_1)^2+{y_1}^2+
2\frac{m_{\Lambda}}{M_1}x_1y_1v_1)^{1/2},
$$
$$
x'_2=((\frac{\mu_2}{m_{\Lambda}}x_2)^2+{y_2}^2+2\frac{\mu_2}{m_{\Lambda}}x_2y_2v_2)^{1/2},
\quad y'_2=\frac{1}{2}((\frac{m_{\Lambda}}{M_2}x_2)^2+{y_2}^2-
2\frac{m_{\Lambda}}{M_2}x_2y_2v_2)^{1/2},
$$
$$
x''_2=((\frac{\mu_2}{m_C}x_2)^2+{y_2}^2-2\frac{\mu_2}{m_C}x_2y_2v_2)^{1/2},
\quad y''_2=\frac{m_{\Lambda}}{m_C+m_{\Lambda}}((\frac{m_C}{M_2}x_2)^2+{y_2}^2
+2\frac{m_C}{M_2}x_2y_2v_2)^{1/2}.
$$
Here $v_j$ is the cosine of angle between ${\bf x_j}$ and ${\bf y_j}$. 
The $s$-wave total wavefunction then assumes the form 
\begin{equation} 
\label{eq:psi} 
\Psi(x_1,y_1,v_1)={\cal U}(x_1,y_1)/(x_1y_1)+A{\cal W}(x'_1,y'_1)/(x'_1y'_1)
+A{\cal W}(x''_1,y''_1)/(x''_1y''_1) \;\;, 
\end{equation} 
where the product $A{\cal W}$ ensures that the two $\Lambda$'s are in 
a $^1S_0$ state as required by the Pauli principle. 
Note that the squares of elements of $A$ correspond then to 
a $(2J+1)$ average over the $J = 0,1$ (singlet and triplet, respectively) 
$^4_\Lambda$H and $^4_\Lambda$He states. We also note in passing that the 
$s$-wave Faddeev equations for $^{~~6}_{\Lambda\Lambda}$He 
($\alpha\Lambda\Lambda$) and for $^{9}_{\Lambda}$Be ($\Lambda\alpha\alpha$) 
are obtained from Eqs. (\ref{eq:fad}) by setting $A=B=1$ with obvious 
changes in the notation for the masses and the total wavefunction (and adding 
the Coulomb potential between the two $\alpha$'s in $^{9}_{\Lambda}$Be). 

\subsection{Yakubovsky equations}

The four-particle wavefunction $\Psi$ is decomposed, in the 
Faddeev-Yakubovsky method \cite{MFa93}, into components which are 
in a one to one correspondence with all chains of partitions. The 
chains consist of two-cluster partitions $a_2$ ({\it e.g.}, $(ijk)l$ 
or $(ij)(kl)$) and three-cluster partitions 
$a_3$ ({\it e.g.}, $(ij)kl$) obeying the relation 
${a_3}\subset {a_2}$. The latter means that the partition $a_3$ can be 
obtained from the partition $a_2$ by splitting up one subsystem. 
Any three-cluster partition $a_3$ may be obtained in this way from 
three different two-cluster partitions $b_2$, $c_2$ and $d_2$. 
It is easy to see that there exist generally $18$ chains of partitions 
for a four-particle system. The components $\Psi_{{a_3}{a_2}}$ obey 
the differential Yakubovsky equations
\cite{MYG84} 
\begin{equation} 
\label{eq:Ya}
(H_{0}+V_{a_3}-E)\Psi_{{a_3}{a_2}}+V_{a_3}\sum_{({c_3}\neq {a_3})}
\Psi_{{c_3}{a_2}} = 
-V_{a_3}\sum_{{d_2}\neq {a_2}}\sum_{({d_3}\neq {a_3})\subset 
{a_2}}\Psi_{d_{3}d_{2}}\;\;\;. 
\end{equation} 
Here $H_0$ is the kinetic-energy operator and $V_{a_3}$ stands 
for the two-body potential acting within the two-particle subsystem 
of a partition $a_3$ ({\it e.g.}, $V_{a_3}$ = $V_{ij}$ with 
$a_3$ = $(ij)kl$).

We demonstrate the Yakubovsky equations for the case of two pairs, each 
consisting of identical constituents. To be more specific we consider 
the pair (1,2) to consist of two identical fermions, and the pair (3,4) 
to consist of two identical bosons, altogether simulating the four-body 
$\Lambda\Lambda\alpha\alpha$ system. This results in a reduction of the 
number of Faddeev-Yakubovsky components into seven independent components 
satisfying the following Yakubovsky differential equations: 
\begin{equation} 
\label{eq:Yakub1} 
(H_0+V^c+V_{\alpha\alpha}-E)U_1 + V_{\alpha\alpha}(I+P_{34})U_2 = 
- V_{\alpha\alpha}\left( (I+P_{34})U_4 - P_{12}(I+P_{34})W_3 \right), 
\end{equation} 
\begin{equation} 
\label{eq:Yakub2} 
(H_0+V^c+V_{\Lambda\alpha}-E)U_2 + V_{\Lambda\alpha}(U_1+ P_{34}U_2) = 
- V_{\Lambda\alpha}\left( P_{34}U_4-P_{12}U_1+W_2-P_{12}W_3 \right), 
\end{equation} 
\begin{equation} 
\label{eq:Yakub3} 
(H_0+V^c+V_{\Lambda\Lambda}-E)U_3 + V_{\Lambda\Lambda}(I-P_{12})U_4= 
- V_{\Lambda\Lambda} \left( (I-P_{12})U_2+ (I-P_{12})P_{34}W_3 \right), 
\end{equation} 
\begin{equation} 
\label{eq:Yakub4} 
(H_0+V^c+V_{\Lambda\alpha}-E)U_4 + V_{\Lambda\alpha}(U_3-P_{12}U_4)= 
- V_{\Lambda\alpha} \left( P_{34}U_3-P_{12}U_2+ W_1+P_{34}W_3 \right), 
\end{equation} 
\begin{equation} 
\label{eq:Yakub5} 
(H_0+V^c+V_{\Lambda\Lambda}-E)W_1 + V_{\Lambda\Lambda}W_2 = 
- V_{\Lambda\Lambda}(I-P_{12})U_1, 
\end{equation} 
\begin{equation} 
\label{eq:Yakub6} 
(H_0+V^c+V_{\alpha\alpha}-E)W_2 + V_{\alpha\alpha}W_1 = 
- V_{\alpha\alpha}(I+P_{34})U_3, 
\end{equation} 
\begin{equation} 
\label{eq:Yakub7} 
(H_0+V^c+V_{\Lambda\alpha}-E)W_3 - V_{\Lambda\alpha}P_{12}P_{34}W_3 = 
- V_{\Lambda\alpha}(U_4+U_2), 
\end{equation} 
where  $U_\beta$ ($\beta$=1,2,3,4) are Yakubovsky components corresponding 
to three-cluster partitions evolving from 3+1 type breakup into two-cluster 
partitions, $W_\gamma$ ($\gamma$=1,2,3) are Yakubovsky components corresponding 
to three-cluster partitions evolving from 2+2 type breakup into two-cluster 
partitions: 
\begin{equation} 
\label{eq:U} 
U_1=\Psi_{(34)2,1}\;,\;\; U_2=\Psi_{(24)3,1}\;,\;\; 
U_3=\Psi_{(12)4,3}\;,\;\; U_4=\Psi_{(24)1,3}\;, 
\end{equation} 
\begin{equation} 
\label{eq:W} 
W_1=\Psi_{(12),34}\;,\;\; W_2=\Psi_{(34),12}\;,\;\; 
W_3=\Psi_{(13),24}\;, 
\end{equation} 
in the shorthand notation of Ref. \cite{KGl92} for the ($a_3$,$a_2$) 
chains of partitions. For each Yakubovsky component the natural 
Jacobi relative coordinates are employed. For the particular 
$\Lambda\Lambda\alpha\alpha$ problem under consideration, 
$V^c$ stands for the Coulomb interaction between 
particles 3 and 4, expressed in terms of the coresponding Jacobi 
coordinates. There are four rearrangement channels altogether, 
defined in terms of the following wavefunctions: 
\begin{equation} 
\label{eq:Psi1} 
\Psi_{(\alpha\alpha\Lambda)\Lambda} = 
(I-P_{12})U_1+(I+P_{34}-P_{12}P_{34}-P_{12})U_2 \;\;, 
\end{equation} 
\begin{equation} 
\label{eq:Psi2} 
\Psi_{(\alpha\Lambda\Lambda)\alpha} = 
(I+P_{34})U_3+(I+P_{34}-P_{12}P_{34}-P_{12})U_4 \;\;, 
\end{equation} 
\begin{equation} 
\label{eq:Psi34} 
\Psi_{(\alpha\alpha)(\Lambda\Lambda)} = W_1+W_2 \;, \;\; 
\Psi_{(\alpha\Lambda)(\alpha\Lambda)} = (I+P_{34}-P_{12}P_{34}-P_{12})W_3 \;. 
\end{equation} 
The total wavefunction can be written as 
\begin{equation}
\label{eq:Psi} 
\Psi = \Psi_{(\alpha\alpha\Lambda)\Lambda}+\Psi_{(\alpha\Lambda\Lambda)\alpha}
+\Psi_{(\alpha\alpha)(\Lambda\Lambda)}+\Psi_{(\alpha\Lambda)(\alpha\Lambda)}\;. 
\end{equation} 
The probability weights for the various rearrangement channels $\beta$ in the 
total wavefunction of $_{\Lambda\Lambda}^{~10}$Be are given by Eq. (\ref{eq:prob}) 
of the previous subsection.  

\subsection{Computation} 

\begin{figure} 
\epsfig{file=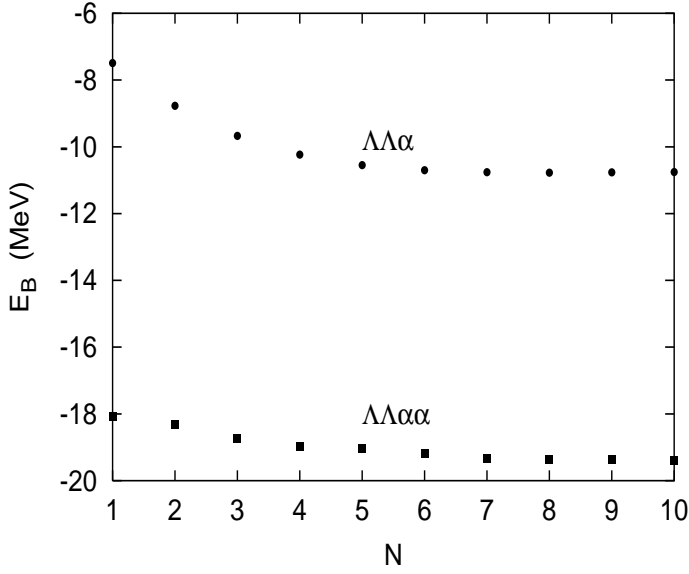, height=75mm,width=90mm} 
\vspace*{5mm} 
\caption{$_{\Lambda\Lambda}^{~~6}$He and 
$_{\Lambda\Lambda}^{~10}$Be ground-state energies $E_B$ 
calculated by solving the $s$-wave Faddeev equations for 
$\alpha\Lambda\Lambda$, and the Yakubovsky equations for 
$\alpha\alpha\Lambda\Lambda$, using model ESC00, 
as function of the number of terms $N$ retained in the 
expansion (\ref{eq:proj}). A cutoff radius 
$R_{\rm cutoff}=19$ fm was used in both calculations.}
\label{fig:conv} 
\end{figure}

The $s$-wave Faddeev-Yakubovsky equations are solved using the cluster 
reduction method (CRM) which was developed by Yakovlev and Filikhin 
\cite{YFi95} and has been recently applied to calculate bound states and 
low-energy scattering for systems of three and four nucleons \cite{FYa99}. 
In this method, the Faddeev-Yakubovsky components are 
decomposed in terms of the eigenfunctions of the Hamiltonians of the 
two- or three-particle subsystems. Due to the projection onto elements 
of an orthogonal basis, one obtains a set of equations with effective 
interactions corresponding to the relative motion of the various clusters. 
Thus, the Faddeev components ${\cal U}_{\alpha}$, $\alpha$=1,2,3 are written 
in the following form: 
\begin{equation} 
\label{eq:proj} 
{\cal U}_{\alpha}(x,y)= \sum\limits_{l=0}^{\infty }
\phi_{\alpha,l}(x)f_{\alpha,l}(y)\;\;\;. 
\end{equation} 
Here, the basic functions $\phi_{\alpha,l}$ are the solutions of 
the two-body Schr\"odinger equation for the subsystem $\alpha$:
\begin{equation} 
\label{eq:sch}  
(-\frac{\hbar^2}{2\mu_{\alpha}}\partial _x^2+v_\alpha(x))\phi_{\alpha,l}(x)=
\varepsilon_{\alpha,l}\phi_{\alpha,l}(x)\;,\;\;\;
\phi_{\alpha,l}(0)=\phi_{\alpha,l}(R_{\rm cutoff})=0\;\;, 
\end{equation} 
where $\mu_{\alpha}$ is the reduced mass of the subsystem $\alpha$, and 
$v_\alpha(x)$ is the $s$-wave projected potential $V_\alpha$. 
Substituting (\ref{eq:proj}) into (\ref{eq:Fad}) and projecting 
onto the orthogonal basis $\phi_{\alpha,l}(x)$ one obtains a set of 
integro-differential equations for the functions $f_{\alpha,l}(y)$. 
A fairly small number of terms is sufficient, using limited computer 
resources \cite{FYa99}, to generate a stable and precise numerical 
solution independently of the value adopted for $R_{\rm cutoff}$, 
provided the latter was chosen sufficiently large. The convergence 
of a typical $_{\Lambda\Lambda}^{~~6}$He Faddeev calculation, 
and of a typical $_{\Lambda\Lambda}^{~10}$Be Yakubovsky calculation, 
as function of the number of terms $N$ retained in Eq. (\ref{eq:proj}) 
above is demonstrated in Fig. \ref{fig:conv}. Such calculations usually 
require less than 10 terms to reach good convergence.
The CRM has been recently used by one of us to study $_{\Lambda}^{3}$H 
as a three-body $pn\Lambda$ system \cite{FYa00a}, and $_{\Lambda}^9$Be 
and $_{\Lambda\Lambda}^{~~6}$He in terms of three-cluster systems 
$\Lambda\alpha\alpha$ and $\alpha\Lambda\Lambda$, respectively \cite{FYa00b}. 
Here we update and extend that calculation, while applying the CRM for 
the first time to the solution of the $\alpha\alpha\Lambda\Lambda$ 
four-body problem in terms of Yakubovsky equations.    

\subsection{Potentials} 

\begin{figure} 
\epsfig{file=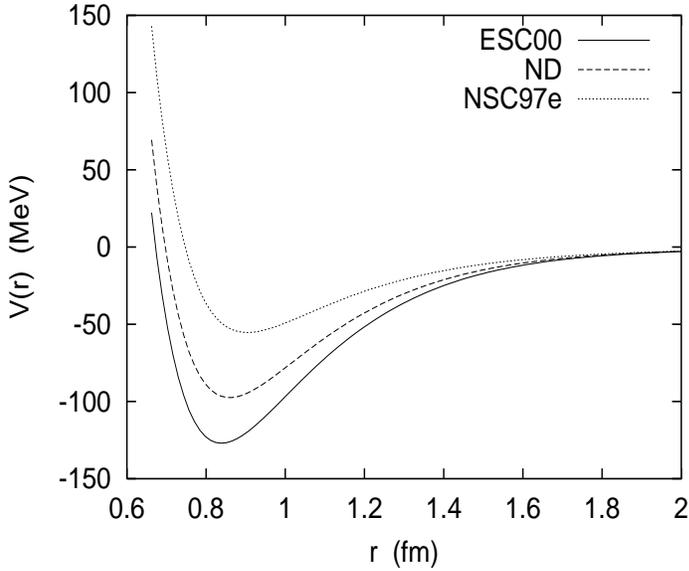, height=75mm,width=90mm} 
\vspace*{5mm}
\caption{$\Lambda\Lambda$ potentials for various simulations of 
Nijmegen models.} 
\label{fig:llpot} 
\end{figure}

The $\Lambda\Lambda$ interaction potentials in the $^{1}S_0$ 
channel which are used as input to the Faddeev equations are of the 
three-range Gaussian form 
\begin{equation} 
\label{eq:HKM}
V_{\Lambda\Lambda} = \sum_i^3 v^{(i)}\exp(-r^2/\beta_i^2)\;\;, 
\end{equation} 
following the work of Hiyama {\it et al.} \cite{HKM97} where 
a phase-equivalent $\Lambda\Lambda$ potential of this soft-core 
form was fitted to the Nijmegen model D (ND) hard-core interaction 
\cite {Nag75} assuming the same hard core for the $NN$ and 
$\Lambda\Lambda$ potentials in the $^{1}S_0$ channel. Our simulation 
of the ND interaction potential is specified, for $\gamma = 1$, 
in Table \ref{tab:tabl0}. For other interactions we have renormalized 
the strength of the medium-range attractive component ($i=2$) of this 
potential such that, using values of $\gamma \neq 1$, it yields 
values for the scattering length and for the effective range as close 
to values prescribed by us. In Table \ref{tab:tabl1} we list the 
low-energy parameters for several Nijmegen potentials, particularly 
the soft-core NSC97 model $\Lambda\Lambda$ potentials \cite{SRi99}, 
and in parentheses the corresponding values for potentials of the 
form (\ref{eq:HKM}) using the listed values of $\gamma$. 
Three of these fitted $^{1}S_0$ potentials are shown in 
Fig. \ref{fig:llpot}. The strongest $\Lambda\Lambda$ potential 
is the one with the largest absolute value of scattering length, 
due to the ESC00 model \cite{Rij01}, and the weakest one is 
associated with the smallest absolute value of scattering length, 
due to one of the NSC97 model potentials. The difference between 
the various NSC97 potentials $(a)$-$(f)$ is of second order, 
all of them being only fairly weakly attractive. 

The $\alpha\alpha$ short-range interaction and the $\Lambda\alpha$ 
interaction are given in terms of two-range Gaussian (Isle) potentials 
\begin{equation} 
\label{eq:Isle}
V^{(j)}=V_{\rm rep}^{(j)}\exp (-r^2/{\beta_{\rm rep}^{(j)}}^2)-
V_{\rm att}^{(j)}\exp (-r^2/{\beta_{\rm att}^{(j)}}^2)\;\;,
\end{equation} 
where the superscript $j$ specifies the interacting system. For the 
$\alpha\alpha$ short-range interaction potential we used the $s$-wave 
component of the Ali-Bodmer potential \cite{ABo66}. 
A Comment about the validity of using this phenomenological potential, 
which does not reflect at short $\alpha\alpha$ distance the underlying 
inner structure of the $\alpha$ clusters, is made below 
in Sect. \ref{sec:res}. The $\Lambda\alpha$ 
interaction potential is due to Ref. \cite{KOA95} where the 
binding energy and mesonic weak decay of $^5_\Lambda$He were studied. 
The resulting $\Lambda$-hyperon density distribution has been shown 
very recently \cite{NAS02} to closely resemble that due to a microscopic 
calculation of $^5_\Lambda$He using $YN$ interactions which simulate 
those of model NSC97 \cite{RSY99}. 
The parameters of these potentials are listed in Table \ref{tab:tabl00}. 
Similar potentials were constructed by us for the $\Lambda - ^3$H and 
$\Lambda - ^3$He singlet and triplet interactions by refitting the range 
parameter $\beta_{\rm rep}$ of the $\Lambda\alpha$ Isle potential 
(\ref{eq:Isle}) to reproduce the known binding energies of $^4_\Lambda$H 
and $^4_\Lambda$He, respectively. These fitted range parameters and the 
binding energies calculated using this procedure are shown in 
Table \ref{tab:tabl2}, in comparison with the experimentally 
known values \cite{DPn86,Bed79}. The $_{\Lambda}^9$Be calculation 
is discussed below in the next section.

\section{Results and Discussion} 
\label{sec:res} 

We first applied the $\alpha\alpha$ and $\Lambda\alpha$ 
potentials, specified by Eq. (\ref{eq:Isle}) and in 
Table \ref{tab:tabl00}, to the solution of the $s$-wave Faddeev 
equations for the $\alpha\alpha\Lambda$ system. 
The calculated ground-state binding energy of $^{9}_{\Lambda}$Be 
in this three-body model is given in Table \ref{tab:tabl2} and is 
in close agreement with the measured $B_{\Lambda}$ value, 
with no need for renormalization or for introducing three-body 
interactions. Had we used a purely attractive $\Lambda\alpha$ 
interaction potential such as the single-Gaussian form due to 
Bando {\it et al.} \cite{BIM82}, the calculated ground-state 
binding energy of $^{9}_{\Lambda}$Be would have come out too 
low by about 1 MeV \cite{FYa00b}. On the other hand the 
calculations by Motoba {\it et al.} \cite{MBI85}, using an Isle 
potential (similar to ours) for the $\Lambda\alpha$ interaction, 
overbind $^{9}_{\Lambda}$Be by about 0.9 MeV, part of which must 
arise from going beyond the $s$-wave dominant 
($\ell,\lambda$)=(0,0) component (98.0\% probability weight). 
This overbinding of $^{9}_{\Lambda}$Be was subsequently cured 
phenomenologically by Hiyama {\it et al.} \cite{HKM97} by 
weakening the $p$-wave component of the underlying $\Lambda N$ 
interaction; alternatively it can be cured introducing a repulsive 
$\Lambda \alpha \alpha$ interaction potential to simulate 
a likely contribution due to $\Lambda NN$ three-body forces 
\cite{BUC84}. Since our interest in $^{9}_{\Lambda}$Be in this 
work is mostly for using it as a subsystem input to calculating 
$^{~10}_{\Lambda\Lambda}$Be, we should strive to use a model 
that reproduces as closely as possible the binding energies 
of such subsystems. The significance of this requirement, 
in the framework of a Faddeev calculation for $^{3}_{\Lambda}$H, 
was emphasized in Ref. \cite{CJF97}. Indeed, our $s$-wave Faddeev 
calculation for $^{9}_{\Lambda}$Be nearly reproduces the observed 
binding energy. The effect of the higher partial waves 
($\ell,\lambda$)=(2,2),(4,4) ignored by us must be largely 
cancelled out in the evaluation of 
$\Delta B_{\Lambda\Lambda}(^{~10}_{\Lambda\Lambda}$Be) 
as defined later on. 

Having thus gained confidence in the appropriateness of the 
$\alpha\alpha$ and $\Lambda\alpha$ input potentials, we then applied 
these potentials to the solution of the Faddeev-Yakubovsky equations 
for {$^{~~6}_{\Lambda\Lambda}$He and $^{~10}_{\Lambda\Lambda}$Be}, 
using different $\Lambda\Lambda$ interactions generically of the 
form (\ref{eq:HKM}). 

\subsection{$^{~~6}_{\Lambda\Lambda}$He and $^{~10}_{\Lambda\Lambda}$Be}

The ground-state energies $E_{\Lambda\Lambda}$ obtained 
by solving the $s$-wave three-body $(\alpha\Lambda\Lambda)$  
Faddeev equations for $^{~~6}_{\Lambda\Lambda}$He and the $s$-wave 
four-body $(\alpha\alpha\Lambda\Lambda)$ Yakubovsky equations for 
$^{~10}_{\Lambda\Lambda}$Be are given in Table \ref{tab:tabl3} for 
$\Lambda\Lambda$ potentials ordered according to their degree of 
attraction. The corresponding $\Lambda\Lambda$ binding energies 
$B_{\Lambda\Lambda}$ are given by 
$B_{\Lambda\Lambda} = - E_{\Lambda\Lambda}$ for 
$_{\Lambda\Lambda}^{~~6}$He and 
$B_{\Lambda\Lambda} = - (E_{\Lambda\Lambda} + 0.1$ MeV) 
for $_{\Lambda\Lambda}^{~10}$Be. The strongest $\Lambda\Lambda$ 
attraction is provided by the potential of the uppermost row, 
simulating the very recent ESC00 model \cite{Rij01} which was 
partly motivated by wishing to get a $B_{\Lambda\Lambda}$ value 
for $_{\Lambda\Lambda}^{~~6}$He as close to that reported by 
Prowse \cite{Pro66}; indeed our calculation reproduces it. 
Significantly lower $B_{\Lambda\Lambda}$ values are obtained for 
our simulation of model ND which, however, reproduces well the 
$B_{\Lambda\Lambda}$ value reported for 
$_{\Lambda\Lambda}^{~10}$Be \cite{DDF89}. These calculated 
$B_{\Lambda\Lambda}$ values were obtained using a $\Lambda\Lambda$ 
potential of the form (\ref{eq:HKM}), very similar to but not 
identical with the one (also marked ND) used in the cluster calculation 
\cite{HKM97} which is listed separately in the table. The appropriate 
low-energy parameters, also shown in Table \ref{tab:tabl3}, are very 
close to each other, and the variation thereby induced in the calculated 
$B_{\Lambda \Lambda}(_{\Lambda\Lambda}^{~~6}$He) is merely 0.03 MeV 
within our own $s$-wave calculation. Furthermore, our calculation 
essentially agrees with that of Ref. \cite{HKM97} for 
$^{~~6}_{\Lambda\Lambda}$He, whereas for $_{\Lambda\Lambda}^{~10}$Be 
our calculation provides about 0.5 MeV more binding. 
In view of the restriction of our calculation to $s$ waves, 
its rough agreement with that of Ref. \cite{HKM97} for 
$_{\Lambda\Lambda}^{~10}$Be should be considered satisfactory. 
We note that the variation within the various calculations of Ref. 
\cite{HKM97} for $B_{\Lambda \Lambda}(_{\Lambda\Lambda}^{~10}$Be), 
as function of the microscopically-motivated $\Lambda\alpha$ 
interactions used by these authors, exceeds 0.5 MeV. We feel that 
typical errors due to limiting our calculations to $s$ waves are 
a fraction of 0.5 MeV for $_{\Lambda\Lambda}^{~~6}$He (as discussed in 
Sect. \ref{sec:sum}) and about 0.5 MeV for $_{\Lambda\Lambda}^{~10}$Be. 
In addition to 
the simulation of the {\it bare} $\Lambda\Lambda$ ND interaction 
discussed above, we also listed in Table \ref{tab:tabl3} binding 
energies calculated for a simulation of the $G$-matrix constructed 
in Ref. \cite{YMH94} out of the ND bare interaction. This was done 
for the sole purpose of comparing below our calculation to those of 
Refs. \cite{YNa00,YMH94}; of course, using a $G$-matrix interaction 
rather than a bare interaction within a full-space calculation is 
fraught with double counting and should be avoided as properly 
acknowledged in the subsequent work of Ref. \cite{HKM97}. 
Down the list in Table \ref{tab:tabl3}, the NSC97 models are seen 
to give yet lower $B_{\Lambda\Lambda}$ values, which for 
$_{\Lambda\Lambda}^{~~6}$He are close to the very recent experimental 
report \cite{Tak01}. In between the rows pertaining to model ND and 
those for model NSC97, just for interpolation, we also listed results 
for two intermediate-strength $\Lambda\Lambda$ interaction potentials 
specified by their low-energy parameters. 

In Tables \ref{tab:LLa} and \ref{tab:LLaa} we list information on 
the geometrical sizes of $^{~~6}_{\Lambda\Lambda}$He and 
$^{~10}_{\Lambda\Lambda}$Be, respectively. As expected, the stronger 
the $\Lambda\Lambda$ interaction is -- the smaller are the geometrical 
sizes of these few-body systems. We note, without offering any 
explanation for it, that almost all of our calculated sizes, particularly 
for $^{~10}_{\Lambda\Lambda}$Be, are significantly larger than those 
calculated in Ref. \cite{HKM97} which is the only comprehensive 
calculation accounting for all rearrangement channels to compare with. 
An interesting exception is provided by the r.m.s. distance between the 
two $\alpha$ clusters in $^{~10}_{\Lambda\Lambda}$Be: 
$R_{\alpha\alpha} = 3.4$ fm in both calculations for model ND, in spite 
of the radically different $\alpha\alpha$ potentials used in these 
calculations. With such a large value of $R_{\alpha\alpha}$, the two 
$\alpha$ clusters apparently are little affected by the inner part of 
their potential, which for a more compressed configuration would require 
a better theoretical guidance than provided by the Ali-Bodmer potential 
\cite{ABo66} used by us. This insensitivity to the inner structure of 
the $\alpha$ clusters, when $\Lambda$'s are added, no longer holds when 
nucleons are added; for example, for the $3\alpha$ system, where 
$R_{\alpha\alpha} = 3.0$ fm in typical $^{12}$C 0$^+$ ground-state 
calculations. Thus, Fedorov and Jensen \cite{FJe96}, using the 
Ali-Bodmer $\alpha\alpha$ potential, had to introduce a strongly 
{\it attractive} $3\alpha$ potential in order to nearly reproduce 
the energies of the two lowest 0$^+$ states in $^{12}$C, whereas 
Hiyama {\it et al.} \cite{HKM97}, excluding Pauli-forbidden 
$\alpha\alpha$ states from their $\alpha\alpha$ potential, 
had to introduce a strongly {\it repulsive} $3\alpha$ potential. 

For $^{~~6}_{\Lambda\Lambda}$He we 
also listed in Table \ref{tab:LLa} the probability (\ref{eq:prob}) 
of having it as a two-body $(\alpha\Lambda) - \Lambda$ cluster. 
This probability goes to one as $V_{\Lambda\Lambda}$ approaches zero; 
the structure of $^{~~6}_{\Lambda\Lambda}$He is then well approximated 
by a $^5_{\Lambda}{\rm He} - \Lambda$ cluster, and the r.m.s. distance 
$R_{\Lambda\alpha}$ between one of the $\Lambda$ hyperons and the 
$\alpha$ cluster is only slightly smaller than the value 
$R_{\Lambda\alpha}(^5_{\Lambda}{\rm He})=3.16$ fm. On the other hand, 
the value of the analogous r.m.s. distance between one of the $\Lambda$ 
hyperons and the c.m. of the $\alpha\alpha$ clusters in 
$^{~10}_{\Lambda\Lambda}$Be, $R_{(\alpha\alpha)\Lambda}=3.26$ fm as 
shown in Table \ref{tab:LLaa} for $V_{\Lambda\Lambda} \rightarrow 0$, 
is considerably larger than the value 
$R_{(\alpha\alpha)\Lambda}(^9_{\Lambda}{\rm Be})=2.88$ fm, indicating that 
the two-body clustering $^9_{\Lambda}{\rm Be}-\Lambda$ is far from saturating 
the structure of $^{~10}_{\Lambda\Lambda}$Be in this limit. This point, 
as well as the information given in parentheses in Table \ref{tab:LLaa}, 
will be discussed in more detail towards the end of the subsection. 
A final point in discussing Tables \ref{tab:LLa} and \ref{tab:LLaa} 
concerns the ground-state energies $E_{\Lambda\Lambda}$ calculated for 
the $G$-matrix interaction ND(G) based on the bare ND interaction. 
Our calculated binding energy for $^{~~6}_{\Lambda\Lambda}$He is higher 
by about 0.8 MeV than that calculated in Refs. \cite{YNa00,YMH94}. 
Given the fact that our calculation for the ND interaction comes very 
close within that of the complete calculation of Hiyama {\it et al.} 
\cite{HKM97}, we conclude that the calculations of Refs. \cite{YNa00,YMH94} 
are far from being complete. 
 
\begin{figure} 
\epsfig{file=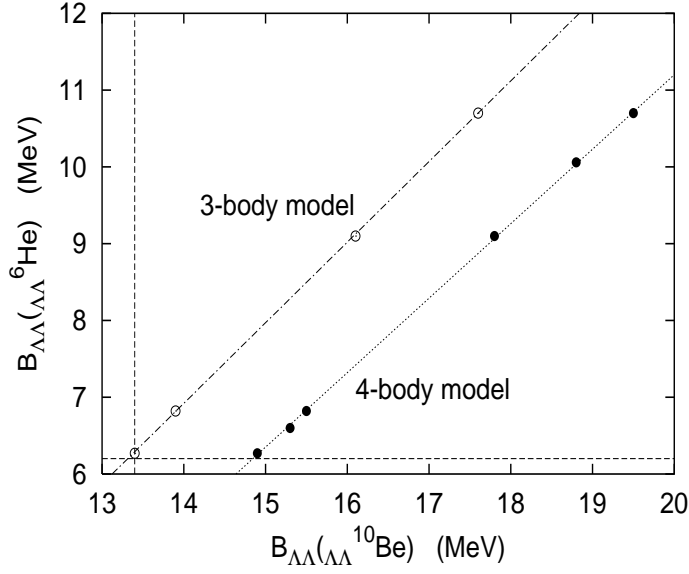, height=75mm,width=90mm}
\vspace*{5mm} 
\caption{Calculated binding energies $B_{\Lambda\Lambda}$ 
for $^{~~6}_{\Lambda\Lambda}$He in a three-body $\alpha\Lambda\Lambda$ 
model, and for $_{\Lambda\Lambda}^{~10}$Be in a four-body 
$\alpha\alpha\Lambda\Lambda$ model and in a three-body 
~$^8$Be $\Lambda \Lambda$ model.} 
\label{fig:e6e10f} 
\end{figure}

It has been shown in previous cluster calculations \cite{BUC84,WTB86} 
that the calculated $B_{\Lambda\Lambda}$ values for 
$^{~~6}_{\Lambda\Lambda}$He and for $^{~10}_{\Lambda\Lambda}$Be are 
correlated nearly linearly with each 
other. Our calculations also indicate such a correlation, as demonstrated 
in Fig. \ref{fig:e6e10f} by the dotted line. This line precludes 
any joint theoretical explanation of the $^{~~6}_{\Lambda\Lambda}$He 
and $^{~10}_{\Lambda\Lambda}$Be experimental candidates listed in Table 
\ref{tab:tabl3}. If $B_{\Lambda\Lambda}(_{\Lambda\Lambda}^{~10}$Be) = 
17.7 $\pm$ 0.4 MeV \cite{Dan63,DDF89}, then the theoretically implied 
$B_{\Lambda\Lambda}$ value for $_{\Lambda\Lambda}^{~~6}$He is about 
9.1 $\pm$ 0.4 MeV, considerably below the value reported by Prowse 
\cite{Pro66} but considerably above the value reported very recently 
by Takahashi {\it et al.} \cite{Tak01}. If 
$B_{\Lambda \Lambda}(_{\Lambda\Lambda}^{~10}$Be) = 14.6 $\pm$ 0.4 MeV, 
on the assumption that the $\pi^-$ weak decay 
of $_{\Lambda\Lambda}^{~10}$Be ground state occurred to the first 
excited doublet levels of $^{9}_{\Lambda}$Be at 3.1 MeV \cite{May83}, 
then the theoretically implied $B_{\Lambda\Lambda}$ value for 
$_{\Lambda\Lambda}^{~~6}$He is 6.1 $\pm$ 0.4 MeV, no longer in such 
a spectacular disagreement with experiment but still significantly 
below the value reported by the recent observation \cite{Tak01} of 
$_{\Lambda\Lambda}^{~~6}$He.

For $V_{\Lambda\Lambda}$ = 0, the lower-left point on the dotted line 
marked `4-body model' in Fig. \ref{fig:e6e10f} corresponds to 
approximately zero incremental binding energy 
$\Delta B_{\Lambda\Lambda}$ for $^{~~6}_{\Lambda\Lambda}$He, where 
\begin{equation} 
\label{eq:delB} 
\Delta B_{\Lambda\Lambda} (^{~A}_{\Lambda \Lambda}Z) 
= B_{\Lambda\Lambda} (^{~A}_{\Lambda \Lambda}Z) 
- 2B_{\Lambda} (^{(A-1)}_{~~\Lambda}Z)\;\;. 
\end{equation} 
This is also explicitly listed in Table \ref{tab:tabl4} and is 
anticipated owing to the rigidity of the $\alpha$ core. However, 
the corresponding $\Delta B_{\Lambda\Lambda}$ value for 
$^{~10}_{\Lambda\Lambda}$Be is fairly substantial, about 1.5 MeV, 
reflecting a basic difference between the four-body 
$\alpha\alpha\Lambda\Lambda$ calculation and any three-body 
approximation in terms of a nuclear core and two $\Lambda$'s as in 
$^{~~6}_{\Lambda\Lambda}$He. To demonstrate this point we show 
by the dot-dash line marked `3-body model' in Fig. \ref{fig:e6e10f} 
the results of a three-body calculation for 
$^{~10}_{\Lambda\Lambda}$Be in which the $^{8}$Be core is not 
assigned an $\alpha\alpha$ structure. In this calculation, the 
$\Lambda-^{8}$Be potential was chosen in a Woods-Saxon form, 
with depth $V_0 = 18.12$ MeV, radius $R = 3.24$ fm and diffusivity 
$a = 0.60$ fm. The geometry and depth were fixed by requiring 
(i) that the ground-state binding energy of the $\Lambda-^{8}$Be 
system agrees with that calculated in the $\alpha\alpha\Lambda$ 
model calculation for $B_{\Lambda}(^{9}_{\Lambda}$Be) 
(see Table \ref{tab:tabl2}); and (ii) that the $\Lambda-^{8}$Be 
r.m.s. distance in the $^{9}_{\Lambda}$Be ground state agrees 
with that between the $\Lambda$ and the c.m. of the two $\alpha$'s 
in the $\alpha\alpha\Lambda$ model calculation for 
$^{9}_{\Lambda}$Be (2.85 fm ${\it vs.}$ 2.88 fm, respectively). 
This three-body calculation for $^{~10}_{\Lambda\Lambda}$Be, 
as is evident from the values given in parentheses in 
Table \ref{tab:LLaa}, gives smaller $B_{\Lambda\Lambda}$ values 
by about 1.7 MeV and also smaller geometrical sizes than those 
given by the four-body calculation. 
The difference is due to the $\alpha\alpha$ correlations which 
are absent in the three-body calculation, and which are built 
in within the Yakubovsky equations of the four-body calculation. 
By breaking up $^{8}$Be into two $\alpha$'s in the four-body 
calculation, substantial attraction is gained due to several 
additional bound subsystems such as for the 
$^{~~6}_{\Lambda\Lambda}$He - $\alpha$ and the 
$^{5}_{\Lambda}$He - $^{5}_{\Lambda}$He clusters which almost 
saturate the corresponding rearrangement channels 
($\alpha\Lambda\Lambda$) - $\alpha$ and 
($\alpha\Lambda$) - ($\alpha\Lambda$), respectively.  
The prominence of these channels is exhibited in 
Table \ref{tab:tabl5} where the probability weights of the various 
rearrangement-channel components of the $^{~10}_{\Lambda\Lambda}$Be 
ground-state wavefunction are listed, using the generic expression 
(\ref{eq:prob}). It is striking how little 
the ($\alpha\alpha$) - ($\Lambda\Lambda$) rearrangement channel 
contributes to the four-body calculation. 
Other calculations \cite{BUC84,WTB86} which pointed out the 
linear relationship discussed above did not account for the full 
range of `attractive' rearrangement channels and consequently 
found smaller values, not exceeding 0.5 MeV, for the 
$\Delta B_{\Lambda\Lambda}(^{~10}_{\Lambda \Lambda}$Be) gain as 
$V_{\Lambda\Lambda} \rightarrow 0$. 

\subsection{The $A=5$ $\Lambda\Lambda$ hypernuclei}

\begin{figure} 
\epsfig{file=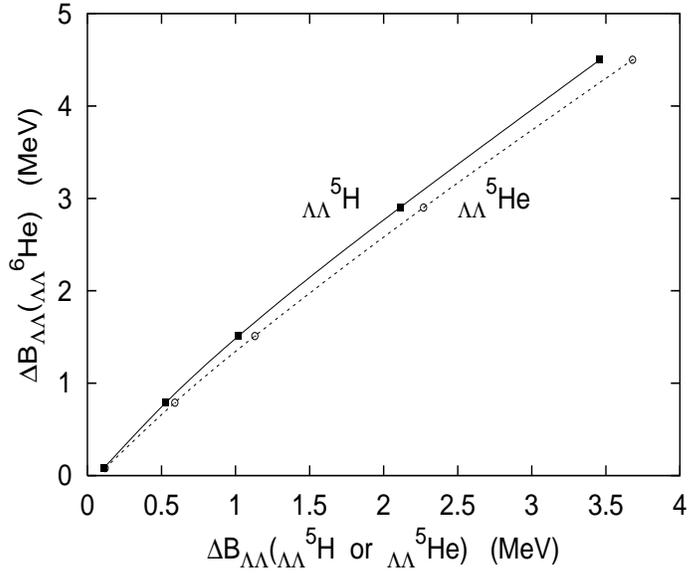, height=75mm,width=90mm} 
\vspace*{5mm} 
\caption{Incremental binding energies $\Delta B_{\Lambda\Lambda}$ 
calculated for $A=5,6$ in a three-body model. 
$\Delta B_{\Lambda\Lambda}$($A$=5) is relative to the (2J+1) average 
of the ($_\Lambda^4$H,~$_\Lambda^4$He) levels of Table \ref{tab:tabl2}.} 
\label{fig:e6e5} 
\end{figure}

The $A = 5$ isodoublet hypernuclei $^{~~5}_{\Lambda\Lambda}$H 
and $^{~~5}_{\Lambda\Lambda}$He were considered by us as three-cluster 
systems $^3$H $\Lambda\Lambda$ and $^3$He $\Lambda\Lambda$, respectively.
The results of Faddeev calculations for these species were included in 
Table \ref{tab:tabl4}, in terms of the incremental binding energy 
$\Delta B_{\Lambda\Lambda}$ (\ref{eq:delB}), together with results 
for $^{~~6}_{\Lambda \Lambda}$He and $^{~10}_{\Lambda \Lambda}$Be.
We note that $\Delta B_{\Lambda\Lambda}$ for $^{~~5}_{\Lambda\Lambda}$He 
is slightly larger than for $^{~~5}_{\Lambda\Lambda}$H, 
owing to the tighter binding of the core $\Lambda$ hypernucleus 
$^{4}_{\Lambda}$He as compared to $^{4}_{\Lambda}$H. 
The $A = 5$ $\Lambda\Lambda$ hypernuclei are found to be particle 
stable for all the $\Lambda\Lambda$ attractive potentials used 
in the present calculation. 
As the $\Lambda\Lambda$ interaction strength goes to zero, 
$\Delta B_{\Lambda\Lambda}(A=5)$ approaches approximately zero, 
similarly to $\Delta B_{\Lambda\Lambda}(A=6)$, provided 
$\Delta B_{\Lambda\Lambda}$($A$=5) is defined {\it with respect to 
the (2J+1) average} of the ($_\Lambda^4$H,~$_\Lambda^4$He) levels 
of Table \ref{tab:tabl2} (this is a physically meaningful 
definition of $\Delta B_{\Lambda\Lambda}$ when several spin-flip 
nuclear core levels contribute to the Faddeev calculation; see also 
the comment made following Eq. (\ref{eq:psi}) for the wavefunction 
of these $A = 5$ $\Lambda\Lambda$ hypernuclei). 
We recall that both of the $A = 5,6$ $\Lambda\Lambda$ hypernuclear 
systems are treated here as three-body clusters. A nearly linear 
correlation appears between $\Delta B_{\Lambda\Lambda}(A=6)$ and 
$\Delta B_{\Lambda\Lambda}(A=5)$, as shown in Fig. \ref{fig:e6e5}. 
Judging by the slope of the straight lines prevailing over most 
of the interval shown in the figure, the $\Lambda\Lambda$ 
interaction is more effective in binding $^{~~6}_{\Lambda\Lambda}$He 
than binding either one of the $A=5$ $\Lambda\Lambda$ hypernuclei. 
For this range of mass values, the heavier the nuclear core is -- 
the larger $\Delta B_{\Lambda\Lambda}$ is (except for a vanishingly 
weak $V_{\Lambda\Lambda}$ where the trend is opposite), 
implying that no saturation is yet reached. 
Within such three-body models $\Delta B_{\Lambda\Lambda}$ appears 
to get saturated already about $A=10$, judging by the slope of the 
three-body calculation straight line of Fig. \ref{fig:e6e10f}. 
However, the special four-body cluster structure of 
$^{~10}_{\Lambda\Lambda}$Be reverses this trend of saturation, 
apparently delaying it into somewhat heavier species 
(for considerably heavier nuclear cores $\Delta B_{\Lambda\Lambda}$ 
should start occasionally decreasing slowly to zero with $A$). 

We have already compared our calculations, discussing Tables 
\ref{tab:tabl3},\ref{tab:LLa},\ref{tab:LLaa}, with those of 
Ref. \cite{HKM97}. In Table \ref{tab:tabl4} we also added 
a comparison with the calculations of Ref. \cite{NSF00}. 
Their $\Delta B_{\Lambda\Lambda}$ values are incredibly higher 
everywhere than ours, and this holds also for their ND-type 
calculation of $_{\Lambda}^5$He.

\section{Summary}
\label{sec:sum}

In this work we studied light $\Lambda\Lambda$ hypernuclear systems 
which may be described in terms of few-cluster systems and treated by 
solving the three-body Faddeev and four-body Faddeev-Yakubovsky 
$s$-wave differential equations. For $^{~10}_{\Lambda \Lambda}$Be, 
the Faddeev-Yakubovsky solution of the $\alpha\alpha\Lambda\Lambda$ 
problem given here is the first one of its kind. We estimate an error 
of about 0.5 MeV due to limiting the $^{~10}_{\Lambda \Lambda}$Be 
calculation to $s$ waves. Our calculation yields more 
binding for $^{~10}_{\Lambda\Lambda}$Be than that produced by most 
of the previous calculations, wherever comparison is meaningful. 
In particular, a fairly large value of $\Delta B_{\Lambda\Lambda} 
\sim 1.5$ MeV survives in the limit $V_{\Lambda\Lambda} \rightarrow 0$ 
owing to the $\alpha\alpha$ structure of the $^8$Be core. Our 
calculation confirms, if not aggravates, the incompatibility 
of the `old' experimental determination of the binding energy of 
$^{~~6}_{\Lambda\Lambda}$He \cite{Pro66} with that of 
$^{~10}_{\Lambda \Lambda}$Be \cite{Dan63} in accordance with 
previous cluster calculations, irrespective of whether or not 
three-body $\Lambda\alpha\alpha$ interactions are included (see 
e.g. Refs. \cite{BUC84,HKM97} respectively). The `new' experimental 
determination of the binding energy of $^{~~6}_{\Lambda \Lambda}$He 
\cite{Tak01} is found to be even more incompatible with that of 
$^{~10}_{\Lambda \Lambda}$Be. Assuming that the determination of 
$B_{\Lambda \Lambda}(^{~10}_{\Lambda \Lambda}{\rm Be})$ 
was plagued by an unobserved $\gamma$ deexcitation 
involving either $^{~10}_{\Lambda \Lambda}$Be$^{*}$ or 
$^{9}_{\Lambda}$Be$^{*}$, somewhere along the sequence of tracks 
observed in the emulsion event carefully reanalyzed in 
Ref. \cite{DDF89}, does not satisfactorily resolve this 
incompatibility. Adding $^{~13}_{\Lambda \Lambda}$B \cite{Aok91} 
as input does not alleviate it either, since the possibility 
of unobserved $\gamma$ deexcitation cannot be dismissed also for 
this species, while on the theoretical side the analysis of 
$^{~13}_{\Lambda \Lambda}$B in terms of a few-cluster system 
is more dubious than for the lighter $\Lambda\Lambda$ species. 
Our conclusion differs radically from the recent claim of 
compatibility by Albertus {\it et al.} \cite{AAN01} who used 
a different methodology, largely untested in studies of very 
light nuclear species. 

Discarding the past history of emulsion experimentation for
$\Lambda \Lambda$ hypernuclear events identified as heavier than
$^{~~6}_{\Lambda \Lambda}$He, because of the ambiguities mentioned
here, one remains with the very recent report from the KEK E373
experiment \cite{Tak01} which claims to have identified uniquely
$^{~~6}_{\Lambda \Lambda}$He, 
with $\Delta B_{\Lambda \Lambda} \sim 1$ MeV. 
No particle-stable excited states are possible for this species
or for its $\Lambda$ hypernuclear core $^{5}_{\Lambda}$He, so this 
event - if confirmed by adding more events of its kind - should be 
taken as the most directly relevant constraint on the 
$\Lambda \Lambda$ interaction. Moreover, $^{~~6}_{\Lambda \Lambda}$He 
is ideally suited for three-body cluster calculations such as the 
Faddeev equations here solved for the $\alpha \Lambda \Lambda$ system. 
Using $s$-wave soft-core $\Lambda \Lambda$ potentials that simulate 
several of the Nijmegen $\Lambda \Lambda$ interaction models, 
we have shown that such simulations of model NSC97, versions $e$ 
and $f$ of which have been shown recently to agree quantitatively 
with data on light $\Lambda$ hypernuclei \cite{RSY99,AHS00}, 
are capable of nearly reproducing the new value \cite{Tak01} of 
$\Delta B_{\Lambda\Lambda}(^{~~6}_{\Lambda\Lambda}$He), short by 
about 0.5 MeV of the mean value.\footnote{in order to precisely 
reproduce this mean value, the parameter $\gamma$ needs to be 
increased from its values listed in Table \ref{tab:tabl1} for 
model NSC97 to 0.6598, yielding the following low-energy 
parameters: $a_{\Lambda \Lambda} = -0.77$ fm, 
$r_{\Lambda \Lambda} = 6.59$ fm.} 
In fact, we estimate the theoretical uncertainty 
of our Faddeev calculation for $^{~~6}_{\Lambda \Lambda}$He 
as a fraction of 0.5 MeV. Two approximations give rise to this 
theoretical uncertainty, 
one is the restriction to $s$-waves in the partial-wave expansion 
of the Faddeev equations, excluding higher $\ell$ values; 
the other one is ignoring the off-diagonal $\Lambda\Lambda - \Xi N$ 
interaction which admixes $\Xi$ components into the 
$^{~6}_{\Lambda \Lambda}$He wavefunction. The following estimates 
may be given for the errors incurred in these approximations. 
As for the $s$-wave approximation, the calculations by Ikeda 
{\it et al.} \cite{IBM85} give 99.7\% probability weight for the 
($\ell,\lambda$)=(0,0) channel, with higher partial waves adding only 
0.3 MeV binding energy, well within the error bars of the measured 
energies. As for including explicitly the $\Lambda\Lambda - \Xi N$ 
coupling, a recent work by Yamada and Nakamoto \cite{YNa00} using 
model ND(G) (another model) finds an increase of 0.4 (0.1) MeV in the 
calculated $B_{\Lambda \Lambda}(^{~~6}_{\Lambda\Lambda}$He) value due 
to including explicitly the $\Lambda\Lambda - \Xi N$ coupling, 
resulting in a 0.3\% (0.1\%) probability $\Xi$ component. 
Carr {\it et al.} \cite{CAG97} found a similar increase of 0.2 MeV 
due to explicitly including this coupling, for a $\Lambda\Lambda$ 
interaction comparable to those used in the present work. 
However, these authors pointed out that, with respect to a calculation 
using an {\it effective} single-channel $\Lambda\Lambda$ interaction 
which implicitly includes the effect of $\Lambda\Lambda - \Xi N$ 
coupling in free space, an explicit two-channel calculation for 
$^{~~6}_{\Lambda\Lambda}$He yields a reduction of 
$B_{\Lambda \Lambda}(^{~~6}_{\Lambda\Lambda}$He) by about 0.3 MeV. 
This reduction is expected due to the tight binding of the $\alpha$ 
cluster which inhibits the effectiveness of the $\Lambda\Lambda - \Xi N$ 
coupling. Therefore, the effects of the two approximations here 
discussed tend to cancel each other almost completely. This conclusion 
should also hold qualitatively for $^{~10}_{\Lambda\Lambda}$Be due 
to the fairly large energy which the $\Lambda\Lambda - \Xi N$ coupling 
requires in order to break an $\alpha$ constituent.    

If model NSC97, as our few-body calculations suggest, provides for the 
right SU(3) extrapolation from fits to $NN$ and $YN$ data, then it 
is questionable whether $^{~~4}_{\Lambda\Lambda}$H is particle stable. 
The subtlties involved in estimating whether or not this species 
provides for the onset of $\Lambda \Lambda$ binding in nuclei were 
clearly demonstrated in Ref. \cite{NMA90}. 
Its $\Delta B_{\Lambda \Lambda}$, at any rate, should not exceed 
0.5 MeV for considerably stronger $\Lambda\Lambda$ interactions 
\cite{NSF00}. If the AGS E906 events conjectured in Ref. \cite{Ahn01} 
as evidence for $^{~~4}_{\Lambda\Lambda}$H are confirmed as such 
in a future extension of this experiment, this four-body system 
$pn \Lambda \Lambda$ would play as a fundamental role for studying 
theoretically the hyperon-hyperon forces as 
$^{3}_{\Lambda}$H $(pn\Lambda)$ has played for studying 
theoretically the hyperon-nucleon forces \cite{MKG95}. 
Our preliminary results, studying the solutions of the 
Faddeev-Yakubovsky equations for the $pn \Lambda \Lambda$ system, 
indicate that $^{~~4}_{\Lambda\Lambda}$H is unstable against 
$\Lambda$ emission. 

\vspace{8mm} 
A.G. would like to acknowledge useful discussions and correspondence 
with Toshio Motoba. This work was partially supported by the Israel 
Science Foundation (grant 131/01). I.N.F. is also partly supported 
by the Russian Ministry of Education (grant E00-3.1-133).

\begin{table} 
\caption{Parameters of the potential (\ref{eq:HKM}) simulating 
${\Lambda\Lambda}$ $^1S_0$ Nijmegen interaction potentials.} 
\label{tab:tabl0} 
\begin{tabular}{ccc} 
$i$ & $\beta_i$ (fm) & $v^{(i)}$ (MeV) \\  \hline 
1 & 1.342 & -21.49 \\ 
2 & 0.777 & -379.1~$\gamma$ \\ 
3 & 0.350 & 9324 \\ 
\end{tabular} 
\end{table}

\begin{table}
\caption{${\Lambda\Lambda}$ $^1S_0$ scattering lengths $a$ and 
effective ranges $r$ for several Nijmegen potential models. 
The values in parentheses correspond to using the form 
(\ref{eq:HKM}), with parameters from Table \ref{tab:tabl0} 
plus the values of $\gamma$ listed here to simulate these potentials.} 
\label{tab:tabl1} 
\begin{tabular}{cccccc}
Parameter & NSC97b & NSC97e & NSC97f & ND & ESC00 \\ \hline 
$a_{\Lambda\Lambda}$ (fm) & -0.38 (-0.38) & -0.50 (-0.50) & 
-0.35 (-0.36) & -2.80 (-2.81) & -10.6 (-10.6) \\
$r_{\Lambda\Lambda}$ (fm) & 10.24 (15.2) & 9.11 (10.6) & 
14.68 (16.6) & 2.81 (2.95) & 2.7 (2.23)  \\ 
$\gamma$ & 0.4804 & 0.5463 & 0.4672 & 1 & 1.2044 \\ 
\end{tabular}
\end{table}

\begin{table} 
\caption{Parameters of the Isle potential (\ref{eq:Isle}) for the 
$\alpha\alpha$ and $\Lambda\alpha$ interactions.} 
\label{tab:tabl00} 
\begin{tabular}{ccccc} 
$j$ & $V_{\rm rep}^{(j)}$ (MeV) & $\beta_{\rm rep}^{(j)}$ (fm) & 
$V_{\rm att}^{(j)}$ (MeV) & $\beta_{\rm att}^{(j)}$ (fm) \\  \hline 
$\alpha\alpha$ \cite{ABo66} & 120.0 & 1.53 & 30.18 & 2.85 \\ 
$\Lambda\alpha$ \cite{KOA95} & 450.4 & 1.25 & 404.9 & 1.41 \\ 
\end{tabular}
\end{table}

\begin{table}
\caption{Binding energies $B_\Lambda$ of $\Lambda$ hypernuclei.}
\label{tab:tabl2}
\begin{tabular}{ccccccc} 
  & $^4_\Lambda$H(0$^+$) & $^4_\Lambda$H(1$^+$) & 
$^4_\Lambda$He(0$^+$) & $^4_\Lambda$He(1$^+$) & $^5_\Lambda$He & 
$^9_{\Lambda}$Be \\ \hline 
$\beta_{\rm rep}$ (fm) & 1.2573 & 1.2720 & 1.2532 & 1.2687 & 1.25 & - \\ 
$B_\Lambda^{\rm calc.}$ (MeV) & 2.06 & 1.04 & 2.36 & 1.24 & 3.09 & 6.64\\ 
$B_\Lambda^{\rm exp.}$ (MeV) & 2.04$\pm$0.04$^a$ & 1.00$\pm$0.04$^b$ & 
2.39$\pm$0.03$^a$ & 1.24$\pm$0.04$^b$ & 3.12$\pm$0.02$^a$ & 
6.71$\pm$0.04$^a$ \\   
\end{tabular}
$^a$Ref. \cite{DPn86}~~~$^b$Ref. \cite{Bed79}
\end{table}

\begin{table}
\caption{Calculated ground-state energies of $\Lambda\Lambda$ hypernuclei 
$E_{\Lambda\Lambda}$ (in MeV) with respect to the breakup threshold of the 
free clusters, using $\alpha\alpha$ and $\Lambda\alpha$ potentials of the 
form (\ref{eq:Isle}) with parameters listed in Table \ref{tab:tabl00}, for 
$\Lambda\Lambda$ potentials (\ref{eq:HKM}) simulated by fitting to the 
scattering length $a$ and effective range $r$ (in fm) of several Nijmegen 
interaction models.} 
\label{tab:tabl3} 
\begin{tabular}{ccccccc} 
Model  & $^{5}_\Lambda$He &  $^9_\Lambda$Be &
$a_{\Lambda\Lambda}$ & $r_{\Lambda\Lambda}$ & 
$^{~~6}_{\Lambda\Lambda}$He & $^{~10}_{\Lambda\Lambda}$Be\\ \hline
ESC00 & -3.09 &  -6.55 & -10.6 & 2.23 & -10.7 & -19.4 \\
ND(G)$^a$       &       &      & -5.37 & 2.40 & -10.1 & -18.7 \\ 
 ND      &       &      & -2.81 & 2.95 & -9.10 & -17.7 \\
 --      &       &        & -0.77 & 2.92 & -7.70 & -16.4 \\ 
 --      &       &        & -0.31 & 3.12 & -6.98 & -15.6 \\ 
NSC97e       &       &        & -0.50 & 10.6 & -6.82 & -15.4 \\
NSC97b       &       &        & -0.38 & 15.2 & -6.60 & -15.2 \\   
$V_{\Lambda\Lambda}=0$   &  &  & 0.0 & -- & -6.27 & -14.8 \\ \hline 
\cite{HKM97} ND  & -3.12 & -6.67 & -2.80 & 2.81 & -9.34 & -17.15\\  
\hline
exp.  & -3.12$\pm$0.02$^b$ & -6.62$\pm$0.04$^b$ & -- & -- & 
-10.9$\pm$0.6$^c$ & -17.6$\pm$0.4$^d$ \\
 & & & & & -7.25$\pm$0.19$^{+0.18}_{-0.11}$~$^e$ & -14.5$\pm$0.4$^f$ \\ 
\end{tabular} 
$^a$Ref. \cite{YMH94}~~~$^b$Ref. \cite{DPn86}~~~$^c$Ref. \cite{Pro66}~~~
$^d$Ref. \cite{DDF89}~~~$^e$Ref. \cite{Tak01}~~~
$^f$assuming ~$^{~10}_{\Lambda\Lambda}$Be ~$\to~ \pi^- ~+~ p ~+~ ^9_\Lambda$Be*
\end{table}

\begin{table} 
\caption{Ground-state energies $E_{\Lambda\Lambda}$ of 
$^{~~6}_{\Lambda\Lambda}$He calculated for an $\alpha\Lambda\Lambda$ system; 
the r.m.s distance between the $\Lambda$ hyperons ($R_{\Lambda\Lambda}$), 
between the center of mass of the hyperon pair to the $\alpha$ cluster 
($R_{(\Lambda\Lambda)\alpha}$) and between one of the $\Lambda$ hyperons 
to the $\alpha$ cluster ($R_{\Lambda\alpha}$); the probability weight 
($P_{(\Lambda\alpha)-\Lambda}$) using Eq. (\ref{eq:prob}) for the 
rearrangement channel $(\Lambda\alpha)-\Lambda$, for various 
$\Lambda\Lambda$ interaction models.} 
\label{tab:LLa} 
\begin{tabular}{cccccc}
Model & $E_{\Lambda\Lambda}$ (MeV) & $R_{\Lambda\Lambda}$ (fm) & 
$R_{(\Lambda\Lambda)\alpha}$ (fm) & $R_{\Lambda\alpha}$ (fm) & 
$P_{(\Lambda\alpha)-\Lambda}$  \\ \hline 
ESC00 & -10.7  & 3.09 & 2.04 & 2.56 & 0.804 \\ 
ND(G) & -10.1  & 3.18 & 2.06 & 2.61 & 0.830 \\ 
ND & -9.10  & 3.36 & 2.11 & 2.70 & 0.870 \\ 
NSC97e & -6.82 & 3.93 & 2.29 & 3.02 & 0.969 \\ 
$V_{\Lambda\Lambda} = 0$ & -6.27 & 4.09 & 2.35 & 3.12 & 0.999 \\  \hline 
\cite{YMH94} ND(G) & -9.23  &  3.31  & 2.14 &  \\ 
\cite{YNa00} ND(G) & -9.4 &  &  &  \\ 
\cite{HKM97} ND & -9.34  &  3.20  &   & 2.55 \\ 
\end{tabular}
\end{table}

\begin{table} 
\caption{Ground-state energies $E_{\Lambda\Lambda}$ of 
$^{~10}_{\Lambda\Lambda}$Be calculated for an $\alpha\alpha\Lambda\Lambda$ 
system; the r.m.s distance between the $\alpha$ clusters 
($R_{\alpha\alpha}$), between the $\Lambda$ hyperons ($R_{\Lambda\Lambda}$), 
between one of the $\Lambda$ hyperons and the c.m. of the $\alpha\alpha$ 
clusters ($R_{(\alpha\alpha)\Lambda}$) and between the c.m. of the 
$\alpha\alpha$ clusters and the c.m. of the $\Lambda$ hyperons 
($R_{(\alpha\alpha)(\Lambda\Lambda)}$), using various 
$\Lambda\Lambda$ interaction models. Results for a three-body 
$^8$Be~$\Lambda\Lambda$ model are given in parentheses.} 
\label{tab:LLaa} 
\begin{tabular}{cccccc}
Model & $E_{\Lambda\Lambda}$ (MeV) & $R_{\alpha\alpha}$ (fm) & 
$R_{\Lambda\Lambda}$ (fm) & $R_{(\alpha\alpha)\Lambda}$ (fm) & 
$R_{(\alpha\alpha)(\Lambda\Lambda)}$ (fm)\\ \hline
ESC00  & -19.4 & 3.3 & 3.7 & 2.87 & 2.2 \\
       & (-17.5) & -- & (3.2) & (2.49) & (1.9) \\
ND(G)  & -18.7 & 3.3 & 3.8 & 2.91 & 2.3 \\ 
       & (-17.0) & -- & (3.3) & (2.52) & (1.9) \\ 
  ND   & -17.7 & 3.4 & 3.9 & 3.02 & 2.3 \\ 
       & (-16.0) & -- & (3.4) & (2.60) & (2.0) \\
NSC97e & -15.4 & 3.5 & 4.2 & 3.17 & 2.4 \\ 
       & (-13.8) & -- & (3.7) & (2.80) & (2.1) \\ 
$V_{\Lambda\Lambda}=0$ & -14.8 & 3.6 & 4.3 & 3.26 & 2.4 \\ 
       & (-13.3) & -- & (3.8) & (2.85) & (2.1) \\ \hline 
\cite{HKM97} ND & -17.15 & 3.40 & 3.02 &  & 1.90  \\
\cite{YMH94} ND(G) & (-17.7) & -- & (2.81) &  & (1.67) \\      
\cite{YNa00} ND(G) & (-17.0) & -- & (2.8) &  &   \\ 
\end{tabular}
\end{table}

\begin{table} 
\caption{Incremental binding energies $\Delta B_{\Lambda\Lambda}$ 
(in MeV) for ${\Lambda\Lambda}$ potentials (\ref{eq:HKM}) specified 
by the scattering length $a$ and effective range $r$ (in fm). 
$\Delta B_{\Lambda\Lambda}$($A$=5) is relative to the (2J+1) average 
of the ($_\Lambda^4$H,~$_\Lambda^4$He) levels of Table \ref{tab:tabl2}.} 
\label{tab:tabl4} 
\begin{tabular}{ccccccc} 
Model & $a_{\Lambda\Lambda}$ & $r_{\Lambda\Lambda}$ & 
$^{~~5}_{\Lambda\Lambda}$H & $^{~~5}_{\Lambda\Lambda}$He & 
$^{~~6}_{\Lambda\Lambda}$He & $^{~10}_{\Lambda\Lambda}$Be\\ \hline 
 ESC00 & -10.6& 2.23 & 3.46 & 3.68 & 4.51 &  6.1\\
   ND  &-2.81& 2.95 & 2.11 & 2.27 & 2.91 &  4.4 \\
   --  &-0.77& 2.92 & 1.02 & 1.13 & 1.51 &  3.1 \\
   --  &-0.31& 3.12 & 0.53 & 0.59 & 0.79 &  2.3 \\
 NSC97e &-0.50& 10.6 & 0.50 & 0.55 & 0.63 & 2.1 \\ 
 NSC97b &-0.38& 15.2 & 0.37 & 0.40 & 0.41 & 1.9 \\ 
 $V_{\Lambda\Lambda}=0$ & 0 & -- & 0.11 & 0.12 & 0.08 & 1.5 \\ \hline 
 \cite{HKM97} ND & -2.80 & 2.81 & -- & -- & 3.10 & 3.74 \\ 
 \cite{NSF00} ND & -2.80 & 2.81 & 2.8 & 2.7 & 4.3 & -- \\ \hline 
exp. &  &  &  &  & 4.7$\pm$0.6$^a$ & 4.3$\pm$0.4$^b$ \\ 
 &  &  &  &  & 1.01$\pm$0.20$^{+0.18}_{-0.11}$~$^c$ & 1.2$\pm$0.4 
 $^d$ \\ 
\end{tabular} 
$^a$Ref. \cite{Pro66}~~~$^b$Ref. \cite{DDF89}~~~$^c$Ref. \cite{Tak01}~~~
$^d$assuming ~$^{~10}_{\Lambda\Lambda}$Be ~$\to~ \pi^- ~+~ p ~+~ ^9_\Lambda$Be* 
\end{table}

\begin{table} 
\caption{Probability weights of rearrangement channels for 
$_{\Lambda\Lambda}^{~10}$Be, evaluated using a generalization of 
Eq. (\ref{eq:prob}) within a four-body $\alpha\alpha\Lambda\Lambda$ 
calculation. In parentheses, the weights for the bound-system 
clusters are given: $^{~~6}_{\Lambda\Lambda}$He for 
$\alpha\Lambda\Lambda$, $^9_{\Lambda}$Be for 
$\alpha\alpha\Lambda$ and $^5_{\Lambda}$He for $\alpha\Lambda$.} 
\label{tab:tabl5} 
\begin{tabular}{ccccc} 
Model & $P_{(\alpha\Lambda\Lambda)-\alpha}$ ~~~ & 
~~~ $P_{(\alpha\alpha\Lambda)-\Lambda}$ ~~~ & 
~~~ $P_{(\alpha\Lambda)-(\alpha\Lambda)}$ ~~~ & 
~~~ $P_{(\alpha\alpha)-(\Lambda\Lambda)}$ \\ \hline 
 ESC00 & 0.450 & 0.314 & 0.227 & 0.0095 \\ 
 & (0.448) & (0.312) & (0.224) &  \\ 
  ND   & 0.379 & 0.359 & 0.253 & 0.0083 \\ 
 & (0.378) & (0.357) & (0.250) &     \\ 
 NSC97e & 0.276 & 0.428 & 0.291 & 0.0048 \\ 
 & (0.276) & (0.424) & (0.287) &     \\ 
 $V_{\Lambda\Lambda}=0$ & 0.260 & 0.445 & 0.290 & 0.0046 \\ 
 & (0.247) & (0.442) & (0.286) &     \\ 
\end{tabular} 
\end{table}

\end{document}